\documentclass[10pt,twocolumn,superscriptaddress,aps,pra,balancelastpage]{revtex4-2}
\usepackage[utf8]{inputenc}
\usepackage{amsmath,amsfonts,amssymb,amsthm,fullpage,color,graphicx,xcolor,physics}
\usepackage[colorlinks]{hyperref}
\usepackage{CJKutf8}

\begin{document}
\begin{CJK}{UTF8}{gbsn}
\title{Measurement and feedforward induced entanglement negativity transition}

\author{Alireza Seif}
\affiliation{Pritzker School of Molecular Engineering, University of Chicago, Chicago, IL 60637}
\author{Yu-Xin Wang (王语馨)}
\affiliation{Pritzker School of Molecular Engineering, University of Chicago, Chicago, IL 60637}
\affiliation{Joint Center for Quantum Information and Computer Science, University of Maryland, College Park, MD 20742, USA}
\author{Ramis Movassagh}
\affiliation{IBM Quantum, MIT-IBM Watson AI Lab,  Cambridge MA, 02142, USA}
\affiliation{Google Quantum A.I., Venice, CA, 90291}
\author{Aashish A. Clerk}
\affiliation{Pritzker School of Molecular Engineering, University of Chicago, Chicago, IL 60637}

\begin{abstract} 
We study the interplay between measurement-induced dynamics and conditional unitary evolution in quantum systems. We numerically and analytically investigate commuting random measurement and feedforward (MFF) processes and find a sharp transition in their ability to generate entanglement negativity as the number of MFF channels varies. We also establish a direct connection between these findings and transitions induced by random dephasing from an environment with broken time-reversal symmetry.  In one variant of the problem, we employ free probability theory to rigorously prove the transition's existence. Furthermore, these MFF processes have dynamic circuit representations that can be experimentally explored on current quantum computing platforms.

\end{abstract}
\maketitle
\end{CJK}

The evolution of quantum systems is influenced differently by measurements versus unitary time evolution. Understanding the dynamics of entanglement under one or both of these two mechanisms is of interest from various perspectives. On one hand, the interplay between these processes can induce entanglement transitions in monitored quantum systems (see e.g., Refs.~\cite{fisher2023random,gullans2020light,ippoliti2021postselection,hoke2023quantum,li2022cross,friedman2022measurement,PhysRevB.101.104301,Diehl2022}). On the other hand, incorporating measurements and adaptive operations into unitary circuits can facilitate the generation of long-range entangled and topologically ordered states, leveraging faster classical communication~\cite{Dennis2002,piroli2021locc,kim2021measurement,verresen2112efficiently,tantivasadakarn2021long,tantivasadakarn2022shortest,tantivasadakarn2023hierarchy,bravyi2022adaptive,PRXQuantum.4.030318}. 
Note that studies of entanglement transitions typically focus on the system state conditioned on measurement outcomes, while in studies of adaptive dynamics, one deterministically prepares entangled states (with a final form that is not contingent on intermediate measurement results).

In this work, we explore a new question at the intersection of these two directions involving measurement-induced dynamics and conditional unitary evolution. Specifically, we
investigate the interplay of multiple commuting random measurement and feedforward (MFF) channels. In our setting, individual random MFF channels are entangling, and as they commute, 
one might assume that this continues to be true even when they are combined.  
The reality is however more complex:  we uncover a distinct transition in their ability to generate entanglement, characterized by negativity~\cite{peres1996separability,horodecki1998mixed,horodecki2001separability}, as we vary the number of MFF channels. Our work stands apart from previous research on disordered open quantum systems, which has primarily focused on their spectral properties~\cite{sa2020spectral,PhysRevLett.123.234103,wang2020hierarchy}. We instead unveil a scenario where this transition clearly emerges as a quantum characteristic of the dynamics.

We explore different variants of this general class of problems. Notably, in one variant, we analytically prove the existence of a sharp transition using tools from free probability. Additionally, these negativity transitions appear to be independent of local degrees of freedom and occur in both spin and bosonic systems. Our findings are also directly linked to a transition in the dynamics of a system coupled to a bath with broken time-reversal symmetry~\cite{Seif2022}. Our work represents one of the rare instances where exact descriptions of entanglement transitions are attainable~\cite{PRXQuantum.4.030333}.

\begin{figure}
    \centering
\includegraphics[width=\columnwidth]{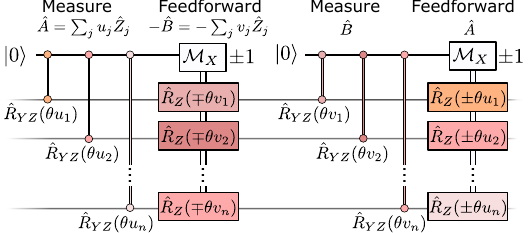}
    \caption{Weak continuous measurements and feedforward of $\hat A$ and $\hat B$ in the $Z$ basis. These operations are can be implemented by repeated coupling to an auxiliary qubit with  $\hat{R}_{YZ}$ rotations proportional to random $u_j$, measuring the qubit, and applying $\hat R_Z$ rotations proportional to random $-v_j$ whose sign is dictated by measurement results ($\pm1$). Here, $\theta$ is a small parameter that determines the timescale.  In the limit of small $\theta$, repeated application of this procedure together with the reverse direction $u_j \to v_j$ and $v_j\to-u_j$ realizes $\mathcal{D}(\hat A + i \hat B)$~\eqref{eq:dissipators}~\cite{supplement,gross2018qubit}.}
    \label{fig:schematic}
\end{figure}

We consider a system of $n$ qubits undergoing continuous measurement and feedforward in the Markovian limit. As we explain below, the unconditional system evolution will follow a Gorini–Kossakowski–Sudarshan–Lindblad~\cite{gorini1976completely,lindblad1976generators} equation $\partial_t \hat{\rho}=\mathcal{L}[\hat\rho]$, where $\hat \rho$ is the system's state at time $t$ and $\mathcal{L}$ is the generator of the dynamics. 

First, let us examine the effect of weak continuous measurement of a Hermitian operator $\hat{A}_k$ on the system. Let $\alpha_k(t)=\langle \hat A _k \rangle + d\xi$ denote the stochastic continuous measurement record, where $d\xi$ is a Wiener increment~\cite{wiseman_milburn_2009}. The trajectories of the system, i.e., the state conditioned on $\alpha_k(t)$, could be entangled, as, in general, non-local measurements can generate entanglement~\cite{vanragemortel2021entanglement,lavasani2022monitored,sriram2022topology}. However, this entanglement is contingent upon the outcome of the measurements; when the record is lost, the entanglement vanishes. Indeed, the dynamics of the unconditional state, i.e., the system's state averaged over the measurement outcomes, is generated by $\mathcal{L}[\hat\rho] = \mathcal{D}(\hat A_k)[\hat{\rho}]$, where
\begin{equation}\label{eq:dissipator_single}
    \mathcal{D}(\hat A_k)[\hat{\rho}] = \hat A_k \hat{\rho} \hat A_k^\dagger -\frac{1}{2} \{\hat A_k^\dagger \hat A_k, \hat \rho\}.
\end{equation} 
In other words, the state is simply being dephased in the measurement basis. 

Next, we add a feed-forward Hamiltonian that takes the measurement signal and drives the system through $-\alpha_k(t) \hat{B}_k$, where $\hat{B}_k$ is some Hermitian system operator. This retains the information about the trajectories in the system and can preserve the entanglement in the unconditional state. As the delay between the measurement and feed-forward operations approaches zero, the dynamics are governed by $\mathcal{L}[\rho]=-i[\hat H_k,\hat \rho]+\mathcal{D}(\hat{A}_k+i\hat{B}_k)[\rho]$, where $\hat H_k=\frac{1}{2}(\hat A_k \hat B_k + \hat B_k \hat A_k)$~\cite{wiseman_milburn_2009,metelmann2017nonreciprocal,supplement}. 

This process can be represented by a dynamic quantum circuit~\cite{ibmdynamics}, where the system is repeatedly and weakly coupled to an auxiliary qubit through $\hat{A}_k$, followed by the measurement of the auxiliary qubit with outcomes $\pm1$. Subsequently, depending on the outcome, we apply a small rotation around $\mp\hat{B}_k$ (see Fig.~\ref{fig:schematic})~\cite{gross2018qubit,supplement}.

In this work, we are interested in studying the simplest possible example of MFF dynamics that demonstrates a transition in entanglement properties of the system. Therefore, we choose commuting $\hat{A}_k$ and $\hat B_k$ that are all diagonal in the energy eigenbasis of the system (implying that the intrinsic system Hamiltonian plays no role). The dynamics might seem trivial with this choice, but as we show, this is surprisingly not the case. Moreover, to avoid breaking reciprocity, we choose bidirectional MFF processes. That is the only symmetry that we impose on the problem.  Therefore, in addition to the weak measurement of $\hat{A}_k$ and driving the system through $-\alpha_k(t) \hat{B}_k$, we also apply the reverse scenario: measuring $\hat{B}_k$ and using the resulting record to apply a drive proportional to $\hat{A}_k$. As a result, the dynamics become fully dissipative with no Hamiltonian component. The directional case exhibits similar physics (see Ref.~\cite{supplement}).  

We aim to explore the general characteristics of entanglement generation with multiple MFF channels. In particular, we are interested in whether the dynamics is entangling, i.e., if there exists some initial product state, such that the evolved state at some future time has non-zero entanglement negativity. Crucially, this entanglement can be transient and is not a steady-state property.   Therefore, we consider the setup introduced earlier with $m$ bidirectional random measurement and feedforward processes, where no additional structure is imposed beyond reciprocity.  We then ask if the dynamics is entangling as we vary the system size $n$ and the number of channels $m$.  To keep the single qubit dephasing rates finite as we vary $m$ and $n$, we normalize the MFF channels  with system size. Specifically, we choose $\hat{A}_k =\frac{1}{\sqrt{n}} \sum_j u_{jk} \hat{Z}_j$ and $\hat{B}_k = \frac{1}{\sqrt{n}}\sum_j v_{jk} \hat{Z}_j$ for $k=1,\dots,m$,  where $\hat{Z}_j$ is the Pauli $\hat \sigma_z$ operator on qubit $j$ and $v_{jk}$ and $u_{jk}$ are chosen independently at  random from a Gaussian distribution
\begin{equation}\label{eq:ginibre}
v_{jk},u_{jk}\sim\mathcal{N}(0,1/2).
\end{equation}
Each channel corresponds to measuring a weighted total spin along the $z$ direction and applying single-qubit rotations proportional to the measurement signal, with different proportionality factors for each spin.
The overall evolution induced by these $m$ channels can be expressed as
\begin{equation}\label{eq:dissipators}
\mathcal{L} = \sum_{k=1}^m\mathcal{D}(\hat A_k + i \hat B_k) = \sum_{k=1}^m \mathcal{D}\left(\frac{1}{\sqrt{n}}\sum_{j=1}^n w_{jk} \hat{Z}_j\right),
\end{equation}
where $w_{jk} = u_{jk}+i v_{jk}$. We can equivalently represent the evolution as
\begin{equation}\label{eq:correlateddephasing}
\partial_t \hat{\rho} = \sum_{i,j=1}^n c_{ij} \left( \hat Z_i \hat\rho \hat Z_j -\frac{1}{2} \{ \hat{Z}_j \hat Z_i, \hat \rho\}\right),
\end{equation}
where $c_{ij}=\frac{1}{n}\sum_{k=1}^m w_{ik}w^*_{jk}$. In matrix notation, this corresponds to $C= \frac{1}{n}WW^\dagger$, where $C = [c_{ij}]\in \mathbb{C}^{n\times n}$ and $W=[w_{ij}]\in \mathbb{C}^{n\times m}$. Thus the correlation between qubits dynamics introduced by various MFF processes can be inferred from $C$. The real part of $C$ contributes to the decay of coherences (off-diagonal elements of $\hat \rho$), while the imaginary part, acting as dissipative Ising-like interactions, results in a phase evolution~\cite{Seif2022}.

\begin{figure}[t]
    \centering
    \includegraphics[width=\columnwidth]{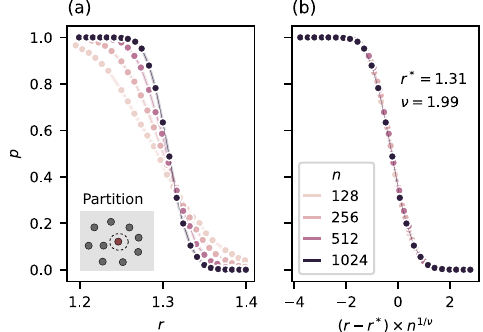}
    \caption{ (a) The fraction $p$ of entangling samples, where each sample corresponds to a realization of $m$ random measurement and feedforward channels in a system of $n$ qubits, and we consider entanglement of one qubit with the remaining $n-1$ qubits.  For large $n$, $p$ undergoes a sharp transition as a function of $r=m/n$. (b) The critical point $r^*$ and the correlation critical exponent $\nu$ are obtained using data collapse. }
    \label{fig:phase_diagram_1vsn}
\end{figure}

To quantify the entanglement, we employ entanglement negativity, which can be computed using the partial-transpose test~\cite{peres1996separability,horodecki1998mixed,horodecki2001separability}. In Ref.~\cite{Seif2022}, it was demonstrated that we can assess the ability of the evolution described in Eq.~\eqref{eq:correlateddephasing} to generate bipartite entanglement negativity between a subsystem $S$ and its complement by examining the spectrum of $\tilde{C}$ obtained from transforming $C$ using the following rule:
\begin{equation}
	\label{eq:pt_hamiltonian}
	\tilde{c}_{ij}=
	\begin{cases}
		-\Re(c_{ij}) & i\in S \text{ and }  j\notin S \text{ (or vice-versa)}\\
		c_{ji} & i\in S \text{ and }  j\in S \\
		c_{ij} & \text{otherwise}
	\end{cases},
\end{equation}
The presence of negative eigenvalues in $\tilde{C}$'s spectrum signifies the capability of the evolution to generate entanglement in the system. This is because the generator of the evolution of the partial transposed state with respect to subsystem $S$ has a similar form to Eq.~\eqref{eq:correlateddephasing} and is given by 
\begin{equation}\label{eq:dissipator_pt}
\begin{split}
     \tilde{\mathcal{L}}_{\rm{diss}}(\hat{\rho}^{T_S})= &- i [\sum_{i,j}\tilde{h}_{ij}\hat{Z}_i \hat{Z}_j,\hat{\rho}^{T_S}]\\&+\sum_{i,j} \tilde{c}_{ij} (\hat{Z}_i \hat{\rho}^{T_S} \hat{Z}_j - \frac{1}{2} \{\hat{Z}_i \hat{Z}_j, \hat{\rho}^{T_S}\}),
\end{split}
\end{equation}
where $\tilde{h}_{ij} = \Im(C_{ij})$ for $i\in S$ and $j\notin S$. In other words, the dissipative Ising interaction $\Im(C_{ij})$ in our original master equation~\eqref{eq:correlateddephasing} plays the role of coherent interactions $\tilde{h}_{ij}$ in the partial-transposed frame. Since the coherent and dissipative parts commute in Eq.~\eqref{eq:dissipator_pt}, we can treat the evolution generated by each independently. The former does not alter the spectrum of  $\hat{\rho}^{T_S}$, while the latter is not necessarily completely positive, and a negative eigenvalue in $\tilde{C}$ indicates the presence of an initially unentangled state ($\hat{\rho}^{T_S}\succeq0$) that becomes entangled ($\hat{\rho}^{T_S}\nsucceq0$) under the evolution described by Eq.~\eqref{eq:correlateddephasing}.

The problem of deciding the entangling power of random MFF channels then boils down to drawing random matrices $W\in\mathbb{C}^{n\times m}$ with varying $n$ and $m$ from a complex Ginibre ensemble~\cite{ginibre1965statistical}, calculating $C=\frac{1}{n}WW^\dagger$, finding $\tilde{C}$ for a given bipartition, and examining $\lambda_{\rm{min}}(\tilde{C})$, the smallest eigenvalue of  $\tilde{C}$. The sign of $\lambda_{\rm{min}}(\tilde{C})$ tells us about the entangling power of $C$. We remark that although our discussion here is focused on qubits, the results have broader applicability; for example, they extend directly to bosonic systems with commuting quadrature Lindblad operators~\cite{supplement}. Moreover, this transition in entangling power of $C$ directly translates to a transition in the negativity of product quantum states orthogonal to the dephasing direction (e.g., $\ket{+}^{\otimes n}$ for qubits and vaccuum for bososns) at a fixed time that has to be short (compared to dephasing rates) for qubits and can be arbitarily long for bosons.  This is because for bosons,  entanglement can increase without bound as purity decreases~\cite{wang2023uncovering}. In contrast, for qubits, entanglement is bounded, and below a certain purity level, mixed state negativity vanishes~\cite{supplement}.

We first numerically study the entanglement negativity between 1 qubit and the rest of the system. Specifically, we examine the probability $p$ of drawing an entangling sample (with $\lambda_{\rm{min}}(\tilde{C})<0$) from the ensemble described above as we vary $r=m/n$~\cite{supplement}. For $m\ll n$, we expect to always have an entangling process ($p=1$) as an individual MFF channel in isolation will in general be entangling~\cite{Seif2022}, and there is a negligible chance for different channels to overlap with each other.  However, in the opposite limit of $m \gg n$, different MFF channels start to overlap, and hence the resulting correlations generated by these processes average away.  This leads to $C \propto I$, implying that the net evolution is equivalent to driving with uncorrelated classical noise, which does not generate entanglement ($p=0$). Therefore, we expect a crossover in $p$ from 0 to 1 as we vary $r$ from 0 to $\infty$. Surprisingly, however, we observe that $p$ goes through a sharp transition from 1 to 0 when $r^*\approx1.3$, i.e., when the number of MFF channels becomes comparable to system size (see Fig.~\ref{fig:phase_diagram_1vsn}a). We also obtain the critical exponent of $\nu\approx2$, by collapsing the data using the scaling form $p=f[(r-r^*)n^{1/\nu}]$ (see Fig.~\ref{fig:phase_diagram_1vsn}b).

Additionally, we analyze the transition using perturbation theory. Let $K = C-\tilde{C}$, where $\tilde{C}$ is obtained from Eq.~\eqref{eq:pt_hamiltonian}. Therefore, the transformation to the partial-transposed frame of Eq.~\eqref{eq:dissipator_pt} can be interpreted as adding a perturbation $K$ to the original coefficients $C$, that is 
\begin{equation}\label{eq:perturbation}
    \tilde{C} = C + K.
\end{equation}
We focus on the asymptotic regime of $n\to\infty$. In this limit, the spectrum of $C$ follows the  Marchenko-Pastur distribution\cite{marchenko1967distribution} 
\begin{equation}\label{eq:marchenkopastur}
    d\mu(\lambda) = 
        \max[0,(1-r)]\delta_0 + \frac{\sqrt{(b-\lambda )(\lambda -a)}}{2\pi \lambda } \mathbf{1}_{[a,b]} d\lambda 
\end{equation}
where $a=(1-\sqrt{r})^2$ and $b=(1+\sqrt{r})^2$. Additionally, $K$ is a rank-2 matrix with eigenvalues $\pm \kappa$, whose magnitude concentrate at $\kappa=\sqrt{r/2}$~\cite{supplement}. 

When $r\ll 1$, $C$ is a low-rank matrix with many 0 eigenvalues. Using degenerate perturbation theory~\cite{sakurai2014modern}, we show that $K$ splits these eigenvalues and $\lambda_{\rm{min}}(\tilde{C})<0$. Degenerate perturbation theory is valid as long as the size of the perturbation ($\sqrt{r/2}$) is smaller than the spacing separating the degenerate subspace from the rest of the spectrum ($(1-\sqrt{r})^2$). Therefore, we have $p=1$ when $\sqrt{r/2}<(1-\sqrt{r})^2$, or equivalently when $r\lessapprox 0.2$. 

In the limit of $r\gg1$, the perturbation is small compared to $\lambda_{\rm{min}}(C)=(1-\sqrt{r})^2$, and therefore cannot change its sign. Specifically, using Weyl's inequality~\cite{weyl1912asymptotische} we have
\begin{equation}\label{eq:weyl}
    \lambda_{\min}({C})+\lambda_{\min}(K)\leq\lambda_{\min}(\tilde{C}).
\end{equation}
Therefore, when $(1-\sqrt{r})^2-\sqrt{r/2}>0$, or equivalently when $r\gtrapprox 5.1$, we have $\lambda_{\min}(\tilde{C})>0$ and consequently $p=0$. 

The constant values of $p=0$ and $p=1$ within these non-vanishing intervals indicate the non-analytic behavior of $p$ as $n\to\infty$. This highlights the critical nature of the observed transition in negativity.

\begin{figure}[t]
    \centering
    \includegraphics[width=\columnwidth]{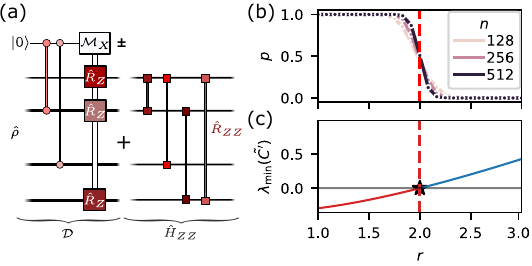}
    \caption{(a) 
    Circuit realization of our modified dynamics, where an additional Ising interaction Hamiltonian $\hat{H}_{ZZ}$ is combined with the MFF channels of the original setup.  $\hat{H}_{ZZ}$ is random but in a way that is correlated with the MFF channels.  This problem has an analytically proven transition. (b) Numerical simulation shows that the transition point is shifted to $r^*=2$. (c) Analytical calculations of $\lambda_{\rm{min}}(\tilde{C}')$ show that its sign changes (and hence the transition happens) at $r^*=2$. }
    \label{fig:phase_independent}
\end{figure}
To go beyond the perturbative treatment and gain a deeper understanding, we must determine the eigenvalues of $\tilde{C}$ in Eq.\eqref{eq:perturbation}. However, computing the eigenvalues of the sum of two matrices is a highly challenging problem that has long captivated mathematicians~\cite{knutson2001honeycombs}. Finding exact solutions to this problem is generally difficult, except in special cases. One remarkable case is that of independent random matrices. In our problem, however, the matrices $K$ and $C$ in Eq.~\eqref{eq:perturbation} are not independent. Nevertheless, we find a slightly different but related physical process by modifying our original problem that allows us to rigorously understand the transition. 

Specifically, we introduce $K'$ as a replacement for $K$, where the elements of $K'$ have the same distribution as $K$ but are now independent of $C$. Consequently, we focus on the alternative problem of finding the smallest eigenvalue of $\tilde{C}' = C + K'$. While this may seem arbitrary and disconnected from the original problem, the assumption of independence has an intriguing physical interpretation: it corresponds to a scenario where, in addition to the dissipative evolution given by Eq.~\eqref{eq:correlateddephasing}, there is an Ising $ZZ$ Hamiltonian (see Fig.~\ref{fig:phase_independent}a) whose coefficients are correlated with the dissipation, given by
\begin{equation}
    \hat{H}_{ZZ} = \sum_{j} [k'_{1j}- \Im(c_{1j})] \hat{Z}_1 \hat{Z}_j .
\end{equation}
This Hamiltonian is entangling. Hence, we expect it to shift the critical point to the right as now there is an additional process contributing to the entanglement generation. This observation is supported by numerical experiments and the following analytical treatment (see Fig.~\ref{fig:phase_independent}b).

To find the spectrum of $\tilde{C}'$ in the large $n$ limit, we use the results of Ref.~\cite{benaych2011eigenvalues} regarding the eigenvalues of low-rank perturbations to large random matrices. The rotational invariance of the eigenvectors of $C$ allows us to find a simple expression for the eigenvalues of $C+\tilde{K'}$. In particular, we find that for $r>2(3+2\sqrt{2})$ the spectrum of $C$ and $\tilde{C}'$ coincide and the perturbation $\tilde{K}'$ does not affect the minimal eigenvalue ($(1-\sqrt{r})^2$), which is consistent with our perturbative analysis~\cite{supplement}. However, for $1<r<2(3+2\sqrt{2})$ the perturbation modifies the spectrum of $C$. In this regime, the minimal eigenvalue of $\tilde{C}'$ is instead given by 
\begin{equation}
    \lambda_{\min}(\tilde{C}')=G^{-1}(\frac{-1}{\sqrt{r/2}}) = r-\frac{3}{\sqrt{2}}\sqrt{r}-\frac{4}{\sqrt{2r}+2}+2, 
\end{equation}
where $G(z)=\int_\mathbb{R} 1/(z-t) d\mu(t)$ is the Cauchy transform of the measure $\mu$~\cite{mingo2017free}. Consequently, we can see that  $\lambda_{\min}(\tilde{C}')<0$ for $1<r<2$, and is non-negative for $r\geq 2$. Hence, $r^*=2$ is the transition point for the entanglement generation in this model (see Figs.~\ref{fig:phase_independent}b and c)~\cite{supplement}. Moreover, using numerical simulations we find that in the modified model $\nu=1.9$ consistent with the original model~\cite{supplement}. While we considered the entanglement negativity of 1 and $n-1$ qubit subsystems here, this analysis can be carried to other finite bipartitions. 

\begin{figure}[t]
    \centering    \includegraphics[width=\columnwidth]{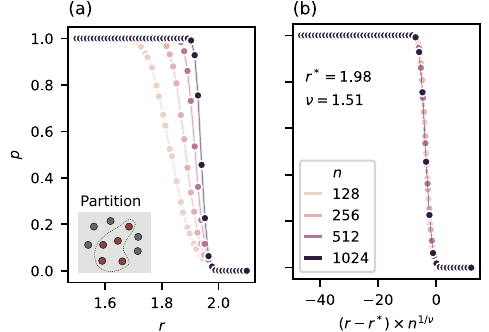}
    \caption{ (a) Entanglement transition similar to Fig.~\ref{fig:phase_diagram_1vsn} for a half system partition. (b) The data collapses with a modified critical point $r^*$ and exponent $\nu$. The shift of the critical point indicates the more robust entanglement in this bipartition.}
    \label{fig:phase_diagram_half}
\end{figure}

The correlated dephasing process in the original model of Eq.~\eqref{eq:correlateddephasing} has several interpretations. So far, we have been interpreting it as the dynamics generated by MFF channels. Alternatively, it could correspond to a general dephasing environment with broken time-reversal symmetry (TRS)~\cite{Seif2022}. Therefore, the question of the entangling power of random MFF channels can be rephrased as 
a quantum-to-classical transition:  
can a random structureless quantum bath with broken TRS generate entanglement or  does it fail to generate entanglement and, in turn, appear as a classical environment from the system's perspective?  
To answer the question about the nature of the environment, we need to check the entangling power for all bipartitions. As we change the size of two partitions from 1 and $n-1$ to $n/2$ and $n/2$ the critical point shifts to the right, i.e., the entanglement is more robust for the half-system bipartition. We repeat the numerical analysis for this case and observe that the sharp transition persists although with shifted $r^*\approx2$ and $\nu\approx1.5$, see Fig.~\ref{fig:phase_diagram_half}.

The observation that $r^*$ remains a constant near 1, even as the partitions become extensive with the system size, suggests that when the number of dephasing channels in the environment becomes comparable to the system size, the system effectively perceives the environment as classical. Thus, even though individual environment channels may possess quantum characteristics, their collective impact does not have any discernible quantum effect on the system. This phenomenon exemplifies a quantum-to-classical transition, where the increasing size of the environment leads to an effective classical behavior. 
Incorporating feedforward into measurement dynamics unveils an intriguing feature in the average post-measurement state of the system, allowing it to retain entanglement generated by measurements. As more MFF channels are introduced, the entangling power of this evolution undergoes a sharp transition. This transition can also be seen as a transition in the nature of the environment as perceived by the system. Remarkably, the existence of this transition is not limited to spin systems, but is, in fact, independent of local degrees of freedom; for example, we observe an analogous transition in bosonic systems with similar kinds of  dynamics~\cite{supplement}.

The MFF channels we discuss in this work can be represented by circuits that are practical for implementation in currently available quantum computers~\cite{baumer2023efficient,iqbal2023topological,foss2023experimental}. Quantum simulations of this entangling transition can shed light on the dynamic interplay between these engineered MFF channels and the inherent noise within the quantum device.

While our study primarily focused on a specific class of random commuting MFF channels, exploring more general cases involving noncommuting MFF channels could yield interesting insights. Additionally, investigating the scaling of entanglement in scenarios with local or sparse interactions (in contrast to the all-to-all connectivity considered in this work) represents a promising avenue for future research.
\begin{acknowledgements}
We thank Tomaz Prosen, Michael Gullans, Yi-Kai Liu, Ali Lavasani, and Tarun Grover for helpful discussions. 

This work was supported by the Air Force Office of Scientific Research MURI program under Grant No. FA9550-19-1-0399, and the Simons Foundation through a Simons Investigator award (Grant No. 669487). AS was partially supported by a Chicago Prize Postdoctoral Fellowship in Theoretical Quantum Science.
\end{acknowledgements}


%

\end{document}


\begin{CJK}{UTF8}{gbsn}
		\title{Supplementary Material for Measurement and feedforward induced entanglement negativity transition}
		\author{Alireza Seif}
		\affiliation{Pritzker School of Molecular Engineering, University of Chicago, Chicago, IL 60637}
		\author{Yu-Xin Wang (王语馨)}
		\affiliation{Pritzker School of Molecular Engineering, University of Chicago, Chicago, IL 60637}
		\affiliation{Joint Center for Quantum Information and Computer Science, University of Maryland, College Park, MD 20742, USA}
		\author{Ramis Movassagh}
		\affiliation{IBM Quantum, MIT-IBM Watson AI Lab,  Cambridge MA, 02142, USA}
		\affiliation{Google Quantum A.I., Venice, CA, 90291}
		\author{Aashish A. Clerk}
		\affiliation{Pritzker School of Molecular Engineering, University of Chicago, Chicago, IL 60637}
		\maketitle
	\end{CJK}

	\appendix
	\renewcommand\thefigure{S\arabic{figure}}

	\section{Negativity transition and the eigenvalue statistics}
	\subsection{Setup\label{sec:setup}}
	We consider an $n$-qubit system, evolving under 
	\begin{equation}\label{eq:lind_sup}
		\partial_t \hat{\rho} = \sum_{i,j=1}^n c_{ij} \left( \hat Z_i \hat\rho \hat Z_j -\frac{1}{2} \{ \hat{Z}_j \hat Z_i, \hat \rho\}\right),
	\end{equation}
	where $C=\frac{1}{n}WW^\dagger$ and  $W\in\mathbb{C}^{n\times m}$ is drawn from a complex Ginibre ensemble~\cite{ginibre1965statistical}. That is, $w_{ik}=u_{ik}+i v_{ik}$, where $u_{ik}$ and $v_{ik}$ are drawn independently at random from $\mathcal{N}(0,\sigma_u)$ and $\mathcal{N}(0,\sigma_v)$, respectively. We vary the system size $n$ and the number of channels $m$. Physically, the parameter $m$ describes the number of random independently chosen measurement and feedforward channels. To decide whether this evolution can generate entanglement negativity for a given bipartition $S$ and $\bar{S}$ we obtain the transformed $\tilde{C}$ by (assume $i\geq j$, while rest of the $\tilde{C}$ matrix can be obtained by $\tilde{C}^\dag=\tilde{C}$)
	\begin{equation}\label{eq:pt_sup}
		\tilde{c}_{ij}=
		\begin{cases}
			-\Re(c_{ij}) & i\in S \text{ and }  j\notin S\\
			c_{ji} & i\in S \text{ and }  j\in S\\
			c_{ij} & \text{otherwise}
		\end{cases},
	\end{equation}
	and examine its minimum eigenvalue denoted by $\lambda_{\rm{min}}(\tilde{C})$. In particular, we are interested in $p$, the probability of having a negative eigenvalue, as we vary $r=m/n$ in the limit of large $m$ and $n$. 
	
	First, following Eq.~\eqref{eq:pt_sup}, we express $\tilde{C}= C+ K$, where $K$ is a hermitian matrix whose elements are given by 
	\begin{equation}
		k_{ij} = \begin{cases}
			- 2\Re(c_{ij})-i\Im(c_{ij}) & \text{if } i=1 \text{ xor } j=1\\
			0 & \text{otherwise}
		\end{cases}.
	\end{equation}
	Note that, however, setting $\Re(K)=0$ does not change the spectrum of $\tilde{C}$. To see this more clearly, let $\{\ket{i}\}_{i=1}^n$ denote a basis for $\mathbb{C}^{n}$. Originally, we have $\tilde{C}=\sum_{ij} \tilde{c}_{ij} \ketbra{i}{j}$, with $\tilde{c}_{ij}\in \mathbb{R}$ if $i=1$ or $j=1$. Therefore, by setting $\Re(K)=0$, we are essentially flipping the sign of  $\tilde{c}_{1j}=\tilde{c}_{j1}$ for $j\neq 1$. This is a unitary transformation that maps $\ket{1}\to - \ket{1}$ and leaves the rest of the states unchanged. Hence, the spectrum of $\tilde{C}$ does not change under this transformation. Therefore, in the following we assume that 
	\begin{equation}
		k_{ij} = \begin{cases}
			-i\Im(c_{ij}) & \text{if } i=1 \text{ or } j=1\\
			0 & \text{otherwise}
		\end{cases}.
	\end{equation}
	Additionally, note that $\rank(K)=2$, as only the first row and columns of $K$ are non-zero. 
	
	To analyze the eigenvalues $\tilde{C}$ we first find the eigenvalues of $C$ and $K$ in Sections~\ref{sec:app_eig_c} and~\ref{sec:app_eig_v}. We then use perturbation theory to examine the eigenvalues of $\tilde{C}$ in small and large $r$ regimes in Section~\ref{sec:app_pert}. Finally, by slightly modifying the original problem we use tools from free probability theory to analytically prove the negativity transition in the modified problem in Section~\ref{sec:app_free}.

	\subsection{Eigenvalue statistics of  $C$ \label{sec:app_eig_c}}
	
	To examine the eigenvalue statistics of $C$ we focus on the case where $u_{ik}$ and $v_{ik}$ have the same distribution and  $\sigma_u=\sigma_v=1/2$. In this case, the density of eigenvalues of a random Wishart matrix $C=\frac{1}{n}WW^\dagger$ as $n,m\to\infty$, while $m/n\to r >0$ follows the Marchenko-Pastur law 
	\begin{equation}
		d\mu(\lambda) = 
		\max[0,(1-r)]\delta_0 + \frac{\sqrt{(b-\lambda)(\lambda-a)}}{2\pi \lambda} \mathbf{1}_{[a,b]} d\lambda,
	\end{equation}
	where $a=(1-\sqrt{r})^2$ and $b=(1+\sqrt{r})^2$. Therefore, the extremal eigenvalues of $C$ are given by 
	\begin{equation}
		\label{eq:lambdaC}
		\begin{split}
			\lambda_{\min}(C) &= \min[0,(1-\sqrt{r})^2], \\ 
			\lambda_{\max}(C) &= (1+\sqrt{r})^2.
		\end{split}
	\end{equation}
	
	Another useful quantity in analyzing the spectrum of random matrices is the the Cauchy transform $G(z)$ of $\mu$ given by~\cite{mingo2017free}
	\begin{equation}
		G(z)=\int_\mathbb{R} \frac{1}{z-x}d\mu(x), 
	\end{equation}
	which is the moment generating function of the eigenvalue density of $C$. We will use it in Section~\ref{sec:app_free} to prove the negativity transition. For simplicity we assume $r>1$ and evaluate 
	\begin{equation}
		G(z)=\int_a^b \frac{\sqrt{(b-x)(x-a)}}{2\pi x(z-x)}dx.
	\end{equation}
	This integral can be evaluated by first changing $x$ to $1+2\sqrt{r}\cos{\theta}+r$ for $\theta\in[0,\pi]$ to obtain 
	\begin{equation}
		G(z) = \int_\pi^0 \frac{2 r \sin^2 \theta  }{\pi  \left(2 \sqrt{r} \cos \theta +r+1\right) \left(2 \sqrt{r} \cos \theta +r-z+1\right)} d\theta, 
	\end{equation}
	which by symmetry can be extended to 
	\begin{equation}
		G(z) = -\int_0^{2\pi} \frac{ r \sin^2 \theta  }{\pi  \left(2 \sqrt{r} \cos \theta +r+1\right) \left(2 \sqrt{r} \cos \theta +r-z+1\right)} d\theta.
	\end{equation}
	We then substitute $w=e^{i\theta}$ and write $G$ as a contour integral 
	\begin{equation}
		G(z) = \frac{1}{4\pi i} \int_\Gamma \frac{r (w-1)^2 (w+1)^2}{w \left(\sqrt{r}+w\right) \left(\sqrt{r} w+1\right) \left(\sqrt{r} w^2+r w+\sqrt{r}-w z+w\right)} dw,
	\end{equation}
	where $\Gamma = \{w\in \mathbb{C} | \abs{w}=1\}$. Only the poles at $w=0,\frac{-1}{\sqrt{r}}$ and $\frac{-\sqrt{(r-z+1)^2-4 r}-r+z-1}{2 \sqrt{r}}$ are inside the contour. Let $f(w)=\frac{r (w-1)^2 (w+1)^2}{4 \pi  i w \left(\sqrt{r}+w\right) \left(\sqrt{r} w+1\right) \left(\sqrt{r} w^2+r w+\sqrt{r}-w z+w\right)}$. We have 
	\begin{align}
		{\rm{Res}}(f,0) &= -\frac{i}{4 \pi } \\
		{\rm{Res}}(f,\frac{-1}{\sqrt{r}}) &= \frac{i \left(\sqrt{r}-1\right) \left(\sqrt{r}+1\right)}{4 \pi  z} \\ 
		{\rm{Res}}(f,\frac{-\sqrt{(r-z+1)^2-4 r}-r+z-1}{2 \sqrt{r}}) &= \frac{i \sqrt{(r-z+1)^2-4 r}}{4 \pi  z}.
	\end{align}
	Therefore, using the residue theorem we find 
	\begin{align}\label{eq:cauchy_sup}
		G(z) &= -\frac{\sqrt{z^2-2 (r+1) z+(r-1)^2}+r-z-1}{2 z}\nonumber\\
		&=\frac{z-r+1-\sqrt{(z-r-1)^2-4r}}{2 z},
	\end{align}
	which is valid for $z\in(\infty,a)\cup(b,\infty)$. 
	
	\subsection{Eigenvalue statistics of $K$\label{sec:app_eig_v}}
	As noted earlier, $K$ has only 2 non-zero eigenvalues. In fact, those eigenvalues are $\lambda_{\pm}=\pm \kappa$, where $\kappa$ is the norm of the first row (or equivalently the first column) of $K$. Specifically, we have 
	\begin{equation}\label{eq:vspec}
		K = \kappa \ketbra{e_+} - \kappa \ketbra{e_-}, 
	\end{equation}
	where $\ket{\pm}$ are the eigenvectors of $K$, and 
	\begin{equation}
		\kappa = \sqrt{\sum_i \Im(c_{1i})^2} = \frac{1}{n}\sqrt{\sum_{i=2}^n \left(\sum_{k=1}^m (u_{1k}v_{ik}-u_{ik}v_{1k})\right)^2}.
	\end{equation}
	In the following we calculate the mean and variance of $\kappa^2$  and argue that in limit of $n\to\infty$, $\kappa$ has a deterministic value. This implies that eigenvalues of $K$ are deterministic. First, we calculate $\expval{\kappa^2}$. We have
	\begin{align}
		\expval{\kappa^2} &= \expval{\sum_{i=2}^n \Im(c_{1i})^2}\\
		&= \frac{1}{n^2}\sum_{i=2}^n\sum_{k_1,k_2 = 1}^m \expval{\prod_{\ell=1}^2 (u_{1 k_\ell}v_{i k_\ell}-u_{i k_\ell}v_{1 k_\ell})}\\
		&= 2 \frac{(n-1)}{n^2} m \sigma_u^2 \sigma_v^2
	\end{align}
	To find the variance of $\kappa^2$, we need to calculate $\expval{\kappa^4}$. We have
	\begin{align}
		\expval{\kappa^4} &= \expval{\sum_{i,j=2}^n \Im(c_{1i})^2\Im(c_{1j})^2}\\
		&=\expval{\sum_{i=2}^n \Im(c_{1i})^4+\sum_{i\neq j}\Im(c_{1i})^2\Im(c_{1j})^2}.
	\end{align}
	For the first term we have 
	\begin{equation}
		\expval{\Im(c_{1i})^4} = \frac{1}{n^4} \sum_{k_1,k_2,k_3,k_4 = 1}^m \expval{\prod_{\ell=1}^4 (u_{1 k_\ell}v_{i k_\ell}-u_{i k_\ell}v_{1 k_\ell})}
	\end{equation}
	Keeping terms with even repeated indices we have two distinct type of terms. First, we have 
	\begin{align}
		\sum_{k_1,k_2,k_3,k_4}\expval{   \prod_{\ell=1}^4 v_{1,k_\ell}u_{i,k_\ell}} &=  m (3\sigma_u^4)(3\sigma_v^4) + 3 m(m-1)\sigma_u^4 \sigma_v^4,
	\end{align} 
	which appears with multiplicity factor of 2. 
	The second term, appearing with multiplicity factor of 6 is
	\begin{align}
		\sum_{k_1,k_2,k_3,k_4}\expval{   \left(\prod_{\ell=1}^2 u_{1,k_\ell}v_{i,k_\ell}\right)\left(\prod_{\ell=3}^4 u_{i,k_\ell}v_{1,k_\ell}\right)} =  m^2 \sigma_u^4 \sigma_v^4 .
	\end{align}
	Therefore, we have 
	\begin{align}
		\expval{\Im(c_{1i})^4} &=\frac{1}{n^4}\{2[m (3\sigma_u^4)(3\sigma_v^4) + 3 m(m-1)\sigma_u^4 \sigma_v^4] + 6 m^2 \sigma_u^4 \sigma_v^4 \}\\
		&= \frac{12}{n^4} m(m+1)\sigma_u^4 \sigma_v^4
	\end{align}
	For the second contribution to $\expval{\kappa^4}$ we need to find 
	\begin{align}
		\expval{\Im(c_{1i})^2\Im(c_{1j})^2} &= \frac{1}{n^4}\sum_{k_1,k_2,k_3,k_4}\expval{\left(\prod_{\ell=1}^2 (u_{1 k_\ell}v_{i k_\ell}-u_{i k_\ell}v_{1 k_\ell}\right)\left(\prod_{\ell=3}^4 (u_{1 k_\ell}v_{j k_\ell}-u_{j k_\ell}v_{1 k_\ell}\right)}.
	\end{align}
	There are only two distinct contributions, each with multiplicity of 2. For the first contribution we have
	\begin{align}
		\sum_{k_1,k_2,k_3,k_4}\expval{\left(\prod_{\ell=1}^4 v_{1,k_\ell}\right)\left(\prod_{\ell=1}^2 u_{i k_\ell}\right)\left(\prod_{\ell=3}^4 u_{j k_\ell}\right)}= 3m \sigma_u^4\sigma_v^4+ m(m-1) \sigma_u^4\sigma_v^4, 
	\end{align}
	and for the second we have
	\begin{align}
		\sum_{k_1,k_2,k_3,k_4}\expval{\left(\prod_{\ell=1}^2 u_{1,k_\ell}v_{i,k_\ell}\right)\left(\prod_{\ell=3}^4 u_{j,k_\ell}v_{1,k_\ell}\right)}
		= m^2 \sigma_u^4 \sigma_v^4. 
	\end{align}
	Therefore, we find 
	\begin{align}    \expval{\Im(c_{1i}^2)\Im(c_{1j}^2)} &= \frac{2}{n^4}(3m \sigma_u^4\sigma_v^4+ m(m-1) \sigma_u^4\sigma_v^4  + m^2 \sigma_u^4 \sigma_v^4 ) \\ 
		&= \frac{4}{n^4}m(m+1)\sigma_u^4 \sigma_v^4 .
	\end{align}
	Finally, we find 
	\begin{align}
		\expval{\kappa^4} &= \frac{1}{n^4}[(n-1)\expval{\Im(c_{1i}^4)} + (n-1)(n-2)\expval{\Im(c_{1i}^2)\Im(c_{1j}^2)}] \\
		&= \frac{1}{n^4}[12(n-1)m(m+1)\sigma_u^4\sigma_v^4 + 4(n-1)(n-2)m(m+1)\sigma_u^4 \sigma_v^4] \\
		&= \frac{4}{n^4} m (m+1)(n^2-1)\sigma_u^4 \sigma_v^4.
	\end{align}
	Therefore, the variance of $\kappa^2$ is given by
	\begin{align}
		\expval{\kappa^4}-\expval{\kappa^2}^2 &= \frac{1}{n^4}\{ 4 m (m+1)(n^2-1)\sigma_u^4 \sigma_v^4 - [2 (n-1) m \sigma_u^2 \sigma_v^2]^2 \}\\
		&= \frac{4}{n^4} m (n-1) (2 m+n+1)\sigma_u^4 \sigma_v^4.
	\end{align}
	We can now expand the mean and standard deviation of $\kappa^2$ at $n\to \infty$ while 
	setting $\sigma_u^2=\sigma_v^2=1/2$ and keeping $r=m/n$ constant to find 
	\begin{align}
		\expval{\kappa^2} &= \frac{r}{2}-\frac{r}{2n}+\mathcal{O}(\frac{1}{n^{2}})\\
		\sqrt{\expval{\kappa^4}-\expval{\kappa^2}^2}&=2 \sqrt{\frac{r (2 r+1)}{2n}} +\mathcal{O}(\frac{1}{n^{3/2}}).
	\end{align}
	Consequently, we see that $\kappa$ concentrates at $\sqrt{r/2}$ when $n\to \infty$. Therefore, in this limit we have 
	\begin{equation}\label{eq:lambdaK}
		\lambda_{\min}(K) = - \lambda_{\max}(K) = -\sqrt{\frac{r}{2}}.
	\end{equation}
	
	\subsection{Perturbative analysis \label{sec:app_pert}}
	\subsubsection{Low-rank $C$ with $r\ll1$}
	First, we examine the case where $r\ll 1$. In this case, $C$ has many 0 eigenvalues as $\rank(C)\leq m$. As we show, using degenerate perturbation theory, $K$ breaks the degeneracy of these 0 eigenvalues and pushes one of them to become negative. 
	
	To find the effect of $K$ on the 0 eigenvalues of $C$, we first project it to this degenerate subspace. Let $\Pi$ denote the projector to $\ker(C)$. We then have to find the eigenvalues of $\Pi K \Pi$. Note that the projection does not increase the rank, and $\Pi K \Pi$ has at most 2 non-zero eigenvalues. Therefore, using Newton's identities, we can relate the eigenvalues of $\Pi K \Pi$ to its first two moments. Specifically, the eigenvalues are the solutions to 
	\begin{equation}
		\lambda^2 - \lambda \Tr(\Pi K \Pi) + \frac{1}{2}[\Tr(\Pi K \Pi)^2-\Tr(\Pi K \Pi K \Pi)] = 0 .
	\end{equation}
	Consequently, the product of the eigenvalues is given by $\omega := \frac{1}{2}[\Tr(\Pi K \Pi)^2-\Tr(\Pi K \Pi K)]$. If $\omega$ is negative, it implies that one of the eigenvalues is negative, and therefore, $C$ is entangling. To find the sign of $\omega$ we use the spectral decomposition of K~\eqref{eq:vspec} and find that 
	\begin{align}
		\omega &= \frac{1}{2}[\Tr(\Pi K \Pi)^2-\Tr(\Pi K\Pi K \Pi)]\\
		&= \frac{1}{2}(\kappa \expval{\Pi}{e_+}-\kappa\expval{\Pi}{e_-})^2-\frac{\kappa^2}{2} 
		[( \langle e_+|\Pi|e_+\rangle ) ^{2}
		+( \langle e_-|\Pi|e_-\rangle ) ^{2}
		- 2| \langle e_+|\Pi|e_-\rangle | ^{2}
		]\\
		&= \kappa^2 [| \langle e_+|\Pi|e_-\rangle | ^{2} - \langle e_+|\Pi|e_+\rangle \langle e_-|\Pi|e_-\rangle- ( \langle e_-|\Pi|e_-\rangle ) ^{2} ].
	\end{align}
	Note that from the Cauchy-Schwarz inequality, we have $| \langle e_+|\Pi|e_-\rangle | ^{2} \leq \langle e_+|\Pi|e_+\rangle \langle e_-|\Pi|e_-\rangle$. Therefore, we conclude that we conclude $ \omega \leq 0$ with equality happening only when $ \Pi|e_-\rangle =0$. In other words, when $K$ does not have any support in $\ker(C)$, or when $K$ has a $1$-dimensional support on $\ker(C)$  corresponding to its positive eigenspace. Neither of these cases are expected to happen unless $r\gtrapprox1$.
	Therefore, $\tilde{C}$ has a negative eigenvalue as long as the perturbation theory is valid and $\Pi K \Pi \not\geq 0$. Degenerate perturbation theory is valid as long as the size of the perturbation ($\kappa=\sqrt{r/2}$) is smaller than the spacing separating the degenerate subspace from the rest of the spectrum ($(1-\sqrt{r})^2$). Therefore, we have $p=1$ when $\sqrt{r/2}<(1-\sqrt{r})^2$, or equivalently when $r\lessapprox 0.2$. 
	
	\subsubsection{Full-rank $C$ with $r \gg 1$}
	When $r\gg 1$, in the limit of $n,m\to\infty$, we can use Weyl's inequality~\cite{weyl1912asymptotische} to bound the minimum eigenvalue of $\tilde{C}$ using the extremal eigenvalues of $C$ and $K$ given by Eqs.~\eqref{eq:lambdaC}~and~\eqref{eq:lambdaK}. Specifically, for the minimal eigenvalue of $\tilde{C}$ we have 
	\begin{equation}
		\lambda_{\min}({C})+\lambda_{\min}(K)\leq\lambda_{\min}(\tilde{C}).
	\end{equation}
	Therefore, we have 
	\begin{equation}
		\lambda_{\min}(\tilde{C})\geq (1-\sqrt{r})^2-\sqrt{\frac{r}{2}}, 
	\end{equation}
	which implies that when $(1-\sqrt{r})^2-\sqrt{r/2}>0$, or equivalently when $r\gtrapprox 5.1$, we have $\lambda_{\min}(\tilde{C})>0$ and consequently $p=0$. 
	
	\subsection{Negativity transition from free probability\label{sec:app_free}} 
	Finding the eigenvalues of $\tilde{C}$ beyond the perturbative limit is challenging as matrices $C$ and $K$ do not commute and are not independent. However, we can find a slightly different but related physical process, in which we can prove the transition. Specifically, we introduce $K'$ as a replacement for $K$, where the elements of $K'$ have the same distribution as $K$ but are now independent of $C$. Consequently, we focus on the alternative problem of finding the smallest eigenvalue of $\tilde{C}' = C + K'$. As noted in the main text this process corresponds to the original process with the addition of a random Ising-type Hamiltonian whose coefficients are correlated with $C$. In this model we can use recent results in free probability to find the spectrum of $\tilde{C}$.
	
	Specifically, let $\tilde{C'} = C + K'$, where $C=\frac{1}{n}WW^\dagger$ is a random Wishart matrix with ordered eigenvalues $\lambda_1(C)\geq\dots\geq\lambda_n(C)$. Similarly, let $\theta_1>0>\theta_{2}$ denote the non-zero eigenvalues of $K'$. Note that in our case $\theta_2 = -\theta_1 = - \sqrt{r/2}$. We then use the following theorem from \cite{benaych2011eigenvalues} to find the minimum eigenvalue of $\tilde {C}$. 
	\begin{thm}
		(Adapted from Theorem 2.1 of Ref.~\cite{benaych2011eigenvalues}) The smallest eigenvalues of $\tilde{C}$
		exhibit the following behavior as $n\rightarrow\infty$
		\[
		\lambda_{n}(\tilde{C})\overset{\text{almost surely}}{\rightarrow}\left\{ \begin{array}{ccc}
			G^{-1}(1/\theta_{2}) & \text{ if } & \theta_{2}<1/G(a^{-})\\
			a &  & \text{otherwise}
		\end{array}\right.
		\]
		where $G(a^{-}):=\lim_{z\uparrow a}G(z)$ and
		$G(z)$ is the Cauchy transform of the eigenvalue density of $C$~\eqref{eq:cauchy_sup},
		and $G^{-1}(z)$ is its functional inverse so that $1/\pm\infty$
		stands for $0$.
	\end{thm}
	\begin{figure}[t]
		\centering
		\includegraphics[width=0.4\textwidth]{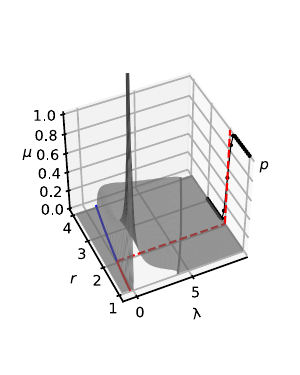}
		\caption{The  distribution $\mu(\lambda)$ of the eigenvalues $(\lambda)$ of $\tilde{C}'$ allows us to analytically find the minimum eigenvalue $\lambda_{\rm{min}}(\tilde{C}')$ (red and blue colored solid line in the $r-\lambda$ plane) and the transition point $r^*=2$. The right plane displays $p$ values as a function of $r$ obtained from numerical simulation of the modified model.  These numerical results indicate that the transition is at $r^*=2$ and are consistent with the analytical findings  (red dashed line).}
		\label{fig:phase_diagram}
	\end{figure}
	Using the results in Eq.~\eqref{eq:cauchy_sup} we find $G(a^-)=\frac{1}{1-\sqrt{r}}$ and  $G(b^+)=\frac{1}{1+\sqrt{r}}$. 
	We also find that the inverse Cauchy transform $G^{-1}(z)$ is given by
	\begin{equation}
		G^{-1}(z) = \frac{(1-r )z-1}{(z-1) z}.
	\end{equation}
	Therefore, we observe that if $\theta_2<1-\sqrt{r}$, the smallest eigenvalue will be given by $G^{-1}(1/\theta_2)$ instead of $a=(1-\sqrt{r})^2$. In our problem, $\theta_2=-\sqrt{r/2}$. Therefore, when $-\sqrt{r/2}<{1-\sqrt{r}}$, or equivalently when $r<2(3+2\sqrt{2})$,  the minimum eigenvalue is given by 
	\begin{equation}
		\lambda_{\min}(\tilde{C})=G^{-1}(\frac{-1}{\sqrt{r/2}}) = r-\frac{3}{\sqrt{2}}\sqrt{r}-\frac{4}{\sqrt{2r}+2}+2, \quad (\text{for } r>1)
	\end{equation}
	The entanglement generation depends on the sign of this quantity. We observe that $\lambda_{\min}(\tilde{C})$ has a root at $r=2$. Therefore, for $1<r<2$, $\lambda_{\min}(\tilde{C})<0$. Hence, $r^*=2$ is the transition point for the entanglement generation. This analytical results fully agree with the numerical simulations as shown in Fig.~\ref{fig:phase_diagram}. Moreover, we perform the scaling analysis and observe that the cricital exponent $\nu=1.9$ consistent with the original model. 
	\begin{figure}[t]
		\centering
		\includegraphics[width=0.5\columnwidth]{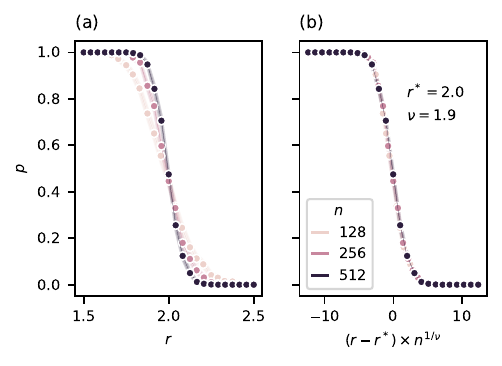}
		\caption{  The modified solvable model. (a) The fraction $p$ of entangling samples, where each sample corresponds to a realization of $m$ random measurement and feedforward channels in a system of $n$ qubits, and we consider entanglement of one qubit with the remaining $n-1$ qubits.  For large $n$, $p$ undergoes a sharp transition as a function of $r=m/n$. (b) The critical point $r^*$ and the correlation critical exponent $\nu$ are obtained using data collapse. }
		\label{fig:phase_diagram_1vsn}
	\end{figure}
	
	\section{Negativity transition with directional measurement and feedforward channels}
	
	In the main text, we considered $m$ bidirectional measurement and feedforward channels defined by operators $\hat{A}_k =\frac{1}{\sqrt{n}} \sum_j u_{jk} \hat{Z}_j$ and $\hat{B}_k = \frac{1}{\sqrt{n}}\sum_j v_{jk} \hat{Z}_j$ for $k=1,\dots,m$. Each
	bidirectional measurement and feedforward channel consists of weak-continuous measurements of $\hat{A}_k$ and a feedforward drive proportional to  $-\hat{B}_k$, and vice versa:  weak-continuous measurements of $\hat{B}_k$ and a feedforward drive proportional to $\hat{A}_k$. However, as mentioned in the main text, in the directional case, where only $\hat{A}_k$ is measured and  a feedforward drive proportional to  $-\hat{B}_k$ is applied, a similar negativity transition occurs.
	
	Specifically, in the directional case, the dissipative part is the same as Eq.~\eqref{eq:lind_sup} with $C=\frac{1}{n}WW^\dagger$, where $W = U + i V$, and $U$ and $V$ are both $n\times m$ matrices whose elements are drawn independently at random from $\mathcal{N}(0,1/2)$. However, in this case, there is also Hamiltonian dynamics generated by 
	\begin{equation}\label{eq:ham_sup}
		\hat{H} = \frac{1}{2}\sum_k \hat{A}_k \hat{B}_k + \hat{B}_k \hat{A}_k = \sum_{i,j} h_{i,j} \hat{Z}_i \hat{Z}_j, 
	\end{equation}
	where $h_{ij} = \frac{1}{2}\sum_k (u_{ik}v_{jk} + u_{jk} v_{ik})$. The existence of this Hamiltonian modifies $\tilde{C}$ so that for a bipartition $S$ and $\bar{S}$, we have $\tilde{c}_{ij}= \tilde{c}_{ji} ^{*}= -\Re(c_{ij}) + i h_{ij}$ for $i\in S$ and $j\notin S$, while leaving the rest of the elements the same as those in Eq.~\eqref{eq:pt_sup}.
	
	\begin{figure}[t]
		\centering
		\includegraphics[width=0.4\textwidth]{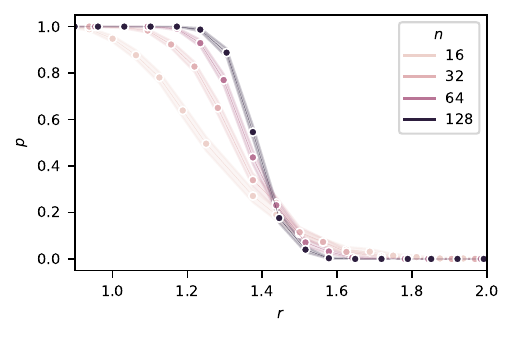}
		\caption{The fraction, $p$, of entangling samples under $m$ directional random measurement and feedforward channels with 1 and $n-1$ partitioning of the system undergoes a sharp transition as a function of the $r=m/n$, where $n$ is the number of qubits.}
		\label{fig:phase_diagram_directional}
	\end{figure}
	
	We again consider a bipartion of 1 and $n-1$ qubits and  using numerical simulations observe that the probability $p$ of drawing an entangling sample (with $\lambda_{\rm{min}}(\tilde{C})<0$) from the ensemble described above as we vary $r=m/n$ undergoes a sharp transition similar to that in Fig.~\ref{fig:phase_diagram_1vsn}. However, the critical point is now slightly moved to the right, possibly due to the addition of the entangling Hamiltonian~\eqref{eq:ham_sup} (see Fig.~\ref{fig:phase_diagram_directional}).  
	
	\section{Negativity transition in quantum states of qubits}
	We consider the state dynamics under
	the master equation~\eqref{eq:lind_sup}. Following Ref.~\cite{Seif2022}, we know that $\hat \rho=|+\rangle\langle+|^{\otimes n}$ is a state that can be used to test the entangling power of the dynamics. The evolution at short times follows 
	\begin{align}
		\hat\rho(dt) & =\hat\rho+dt\sum_{ij}c_{ij}[\hat Z_{i}\hat\rho \hat Z_{j}-\frac{1}{2}(\hat Z_{i}\hat Z_{j}\hat\rho+\hat\rho \hat Z_{i}\hat Z_{j})]\\
		& =\sum_{ij}dtc_{ij}\hat Z_{i}\hat\rho \hat Z_{j}+\hat\rho-dt\hat\rho\sum_{i}c_{ii} -dt(\sum_{i<j}\Re(c_{ij})\hat Z_{i}\hat Z_{j})\hat\rho-dt\hat\rho(\sum_{i<j}\Re(c_{ij})\hat Z_{i}\hat Z_{j})\\
		& =Cdt\oplus\hat\sigma,
	\end{align}
	where $\hat\sigma$ is a positive-semidefinite operator (to first order in $dt$) that depends only on the real part of $C$. The direct sum structure arises from the fact that in short times, the first term in the Lindbladian($\hat Z_{i}\hat\rho \hat Z_{j}$) generates states $\hat{Z_i}\ket{+}^{\otimes n}$, while the other terms generate states in the subspace spanned by $\ket{+}^{\otimes n}$ and $\hat Z_{i}\hat Z_{j} \ket{+}^{\otimes n}$. These two subspaces are orthogonal and therefore we can express $\hat{\rho}(dt)$ as a direct sum of its components in them. 
	
	To take the partial transpose with respect to the first qubit we follow
	our recipe and construct $\tilde{C}$ with elements identical to $C$
	except that $\tilde{c}_{1j}=\tilde{c}_{j1}=-\Re c_{1j}$. We also ignore the $ZZ$ Hamiltonian that emerges in the partial transposed frame as it does not change the spectrum of the partial transposed state~\cite{Seif2022}. Therefore we find 
	\begin{equation}
		\hat\rho^{T_{1}}(dt)=\tilde{C}dt\oplus\tilde{\sigma},
	\end{equation}
	where $\tilde{\sigma}$ is a positive-semidefinite state. The eigenvalues of $\hat\rho^{T_{1}}(dt)$ are the union of eigenvalues of the summands. Therefore, we observe that the negativity of $\hat\rho^{T_{1}}(dt)$ is directly connected to the negative eigenvalues of $\tilde{C}$. While we demonstrated this connection using a single state, there are infinitely many other states that show the same behavior. For example, any tensor product of single qubit states in the $X-Y$ plane of the Bloch sphere have the same dynamics.  
	
\begin{figure}[t]
		\centering
		\includegraphics[width=\textwidth]{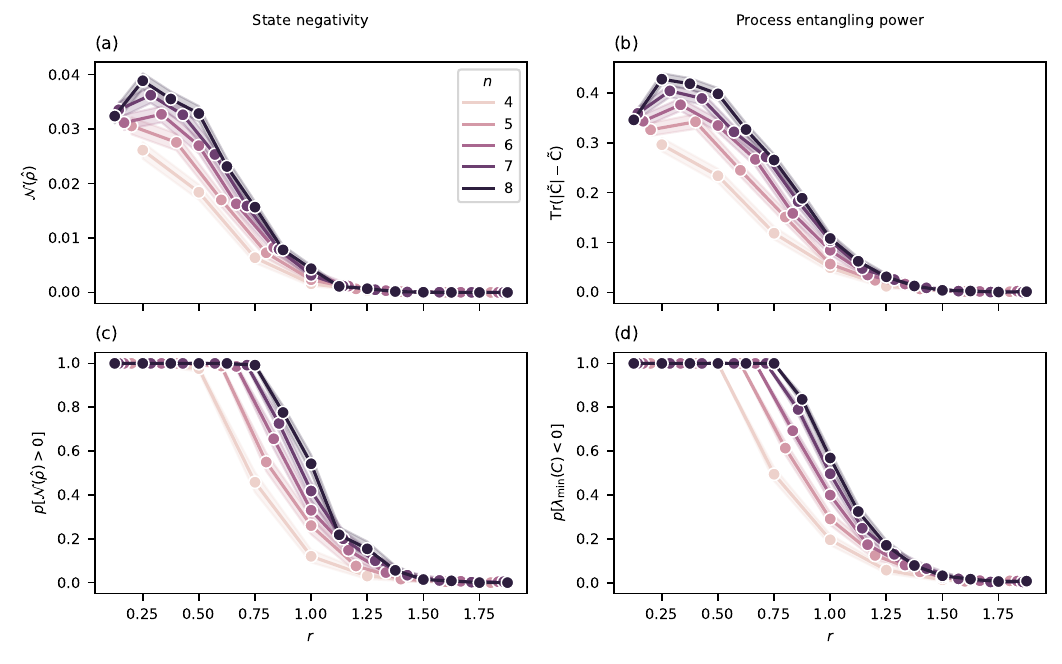}
		\caption{ The relationship between the entangling power of a process and the resulting entanglement negativity in qubits. The candidate state ($\hat{\rho}=\ketbra{+}^{\otimes n}$) is evolved under Eq.~\eqref{eq:lind_sup} to a fixed time $t=0.1$ in units of $C$ for varying $r=m/n$, where $m$ is the number of measurement and feedforward channels and $n$ is the number of qubits. (a) The negativity of the state  $\mathcal{N}(\hat{\rho})$  for various systems sizes $n$, and (b) its process analogue $\tr(\abs{\tilde{C}}-\tilde{C})$ that characterizes the size of negative eigenvalues in $\tilde{C}$, demonstrating a direct translation from the process to the state. (c) The probability $p[\mathcal{N}(\hat{\rho})>0]$ of observing non-zero negativity for the state and (d) the probability $p[\lambda_{\min}(\tilde{C})<0]$ of observing a negative eigenvalue in the process (right). As $r$ is varied the negativity in state or the entangling power of the process change from a non-zero value for small $r$ to 0 for $r\gg1$. The error bars and the shaded region correspond to $95\%$ confidence interval obtained from bootstrapping over 500 samples for the state dynamics and 1000 samples for the process.}
		\label{fig:neg_qubits}
	\end{figure}

	We supplement this discussion by performing a numerical simulation of samples of $n$ qubits evolving under Eq.~\eqref{eq:lind_sup} with random $C$ matrices as prescriped in Sec.~\ref{sec:setup}. We then fix the time $t=0.1$ in units of $C$ and calculate the negativity $\mathcal{N}(\hat\rho) = \frac{1}{2}(\tr(\abs{\hat\rho^{T_A}})-1)$ together with the probability of having a non-zero negativity $p[\mathcal{N}(\hat{\rho})>0]$ over samples. Additionally, we consider analogous quantities for the process and calculate $\tr(\abs{\tilde{C}}-C)$ as well as the probability of $\tilde{C}$ having a negative eignevalue $p[\lambda_{\min}(\tilde{C})<0]$. Our ability to numerically simulate samples of dephasing dynamics with random $C$ matrices in large multiqubit systems is limited, and from these small-scale numerical experiments alone we cannot deduce the existence of a transition. However, the results in Fig.~\ref{fig:neg_qubits} show that indeed the entangling power of the process directly translates to entanglement negativity in the quantum state. Moreover, at a fixed time, these numerical results suggest a continuous transition in negativity as a function of the ratio of the number of channels ($m$) and the number of qubits $r=m/n$. This is expected, as beyond the critical $r^*$ in large enough systems the process does not generate entanglement, whereas before that point the process is entangling and at a fixed time the states will have some entanglement negativity. Turning to bosons in the next section allows us to study the dynamics in much larger systems. 

	\section{Negativity transition in bosonic systems}
	The negativity transition discussed in the main text is not constrained to the qubit-based setup and can be readily generalized to other physical systems. To see this, we first note that Eq.~\eqref{eq:pt_sup} capturing the transformation of master equation under partial transpose is still applicable if we replace any of the qubits in the setup with generic types of systems (e.g., qudits, bosonic modes, or a composite subsystem formed by any of those), and the local qubit operators $\hat Z_j$ in Eq.~\eqref{eq:lind_sup} with any local Hermitian operators acting on corresponding systems. Further, one can prove that in those more general systems, the $\tilde{C}$ matrix having a negative eigenvalue is also a sufficient condition for the dynamics to create entanglement negativity~\cite{wang2023noise}. Thus, the negativity transition we consider and its critical behavior does not rely on the specific choice of the qubit Pauli operators in Eq.~\eqref{eq:lind_sup}.

	Here, we discuss a bosonic version of this transition. Consider a system consisting of $n$ bosonic modes. The setup now takes a similar form as Eq.~\eqref{eq:lind_sup}, where we replace all the local qubit operators $\hat Z_j$ with Hermitian operators acting locally on each bosonic mode. As a concrete example, we focus on dissipative bosonic dynamics that only involves local, mutually commuting quadratures $\hat{x}_j$ ($j=1,2,\ldots,n)$. The system dynamics can be described by the following Lindblad master equation 
	\begin{equation}\label{eq:lind_sup_boson}
		\partial_t \hat{\rho} = \sum_{i,j=1}^n c_{ij} \left( \hat x_i \hat\rho \hat x_j -\frac{1}{2} \{ \hat{x}_j \hat x_i, \hat \rho\}\right),
	\end{equation}
	where $C=\frac{1}{n}WW^\dagger$ and  $W\in\mathbb{C}^{n\times m}$ is again drawn from the complex Ginibre ensemble. 
	As discussed above, in this case we still expect exactly the same negativity transition (in terms of parameter $r\equiv m/n$) as that described in the main text. However, as we now show, the bosonic system in Eq.~\eqref{eq:lind_sup_boson} can exhibit drastically different long-time entanglement dynamics.

\begin{figure}[t]
		\centering
		\includegraphics[width=1\textwidth]{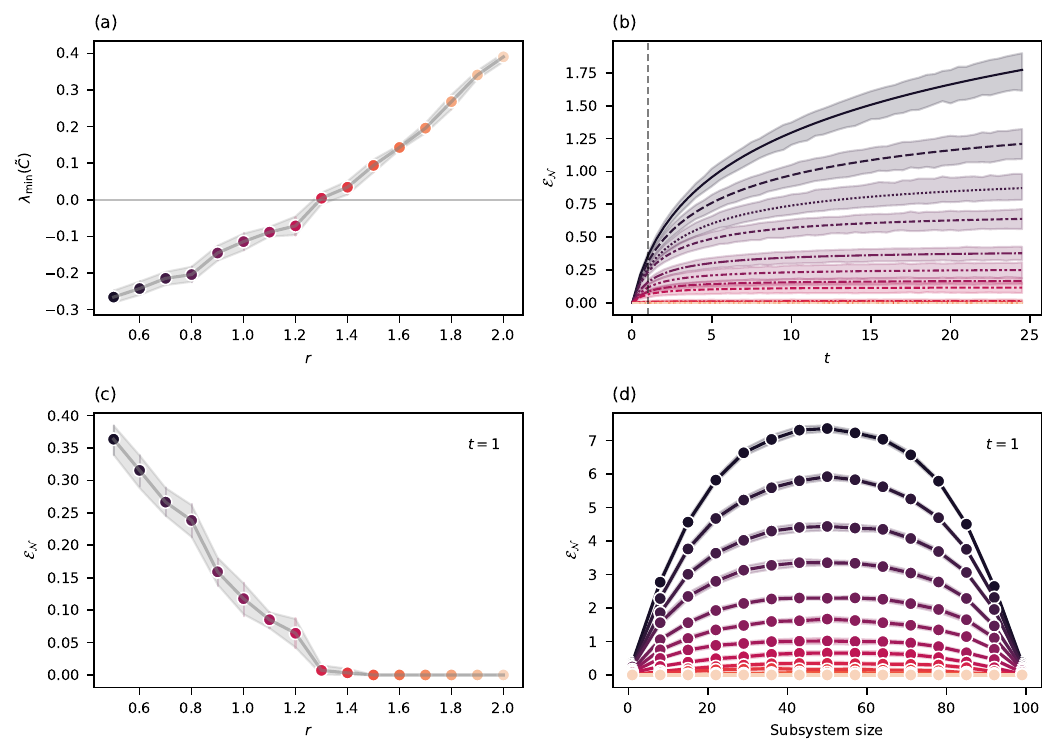}
		\caption{The relationship between the entangling power of a process and the resulting entanglement negativity in bosons for vacuum initial states with $n=100$ bosonic modes.  (a) The minimum eigenvalue of $\tilde{C}$, indicated by $\lambda_{\min}(\tilde{C})$ as a function of $r$.  At $r\approx1.3$, the $\lambda_{\min}(\tilde{C})$ changes sign indicating that beyond this point the process is not entangling. (b) The evolution of log negativity $\mathcal{E}_\mathcal{N}$ of the $(1,n-1)$ partition, as a function of time $t$ for with varying $r$, where the colors of the curves indicate the values of $r$ in panel (a). The curves beyond $r\approx1.3$ remain flat at 0, indicating that there is no negativity generated beyond this point. (c) The log negativity $\mathcal{E}_\mathcal{N}$ as a function of $r$ at a fixed time $t=1$ corresponding to the dashed vertical line in panel (b). Again we observe that beyond the critical point $r\approx1.3$, $\mathcal{E}_\mathcal{N}=0$. Moreover, the results suggest a continuous transition in negativity of the state at a fixed time as a function of $r$. (d) The logarithm negativity $\mathcal{E}_\mathcal{N}$ across different bipartitions (subsystem sizes) at fixed  $t=1$ for different values of $r$. The color of the curves indicates the corresponding values of $r$ in panel (a). We observe that $\mathcal{E}_\mathcal{N}$ grows with subsystem size in the entangling regime, and remains 0 in the non-entangling regime. The error bars and the shaded region correspond to $95\%$ confidence interval obtained from bootstrapping over 10 samples.}
		\label{fig:neg_bosons}
	\end{figure}

	Unlike the qubit case, as Eq.~\eqref{eq:lind_sup_boson} is quadratic in bosonic raising and lowering operators, we can efficiently simulate the full dynamics for much larger systems. Specifically, as the dynamics it generates is Gaussian, time evolution of a Gaussian initial state under Eq.~\eqref{eq:lind_sup_boson} stays Gaussian at any give time $t$. In this case, we can fully characterize the system dynamics $\hat{\rho} (t) $ via only the first two moments of the state. Introducing the vector formed by all quadratures as 
	$\vec{\hat q} \equiv (
	\hat x _{1}, \hat x _{2}, \ldots \hat x _{n},
	\hat p _{1}, \hat p _{2}, \ldots \hat p _{n}
	) $, we now only need to solve all the means 
	$\langle \hat q _{j} \rangle$ and variances
	$ v _{ ij }
	\equiv \langle \{ 
	\delta \hat q _{i}  , 
	\delta \hat q _{j} 
	\} \rangle
	/2$ ($i,j=1,2,\ldots,n$), where $\delta \hat q _{i} 
	\equiv \hat q _{i} - 
	\langle \hat q _{j} \rangle$. Here, $\hat p _{j} $ denote the canonical conjugates of the $\hat x _{j}$ quadratures satisfying $[\hat x _{j} 
	,\hat p _{l} ] = i \delta _{jl} $. As the entanglement properties of a Gaussian state only depend on its variances $v _{ij} (t)$, we now seek to compute the dynamics of those observables. The equations of motion for $v _{ij} (t)$ can be compactly written in terms of a covariance matrix $ V = [ v _{ij}]$, as 
	\begin{align}
		\label{eq:eom_sup_boson}
		\partial _t V
		= M V
		+ V M ^{T} 
		+ F 
		,
	\end{align}
	where $M$ is the dynamical matrix and 
	$F $ describes noise diffusion generated by the dissipative dynamics. We can explicitly write those coefficient matrices as 
	\begin{align}
		M =
		\begin{pmatrix}
			0 & 0  \\
			\Im(C) & 0
		\end{pmatrix} 
		,\, 
		F =\begin{pmatrix}
			0 & 0  \\
			0 & 2  \Re(C)
		\end{pmatrix} 
		.
	\end{align}
	The solution to Eq.~\eqref{eq:eom_sup_boson} directly allows us to compute dynamics of the logarithmic entanglement negativity (see, e.g., \cite{weedbrook2012gaussian}). For a vacuum initial state, the results for states generated by Eq.~\eqref{eq:lind_sup_boson} with random samples of $C$ drawn as prescribed in Sec.~\ref{sec:setup} with different $m$ are shown in Fig.~\ref{fig:neg_bosons}. We see that the entanglement generation for the bosonic case now survives in the asymptotic long-time regime whenever $\lambda_{\rm{min}}(\tilde{C})<0$. Conversely, if $\lambda_{\rm{min}}(\tilde{C})\geq 0$, then Eq.~\eqref{eq:lind_sup_boson} cannot create nonzero negativity from a product initial state. Moreover, we observe that at a fixed time, negativity is non-zero before the critical point at $r\approx 1.3$ and is zero beyond that point, suggesting the existence of a continuous transition. 
	
	\section{Qubit model of measurement and feed forward}
	Here we use the qubit models of weak continuous measurements~\cite{gross2018qubit} and extend them to incorporate feed forward. We consider a scenario that closely parallels conditional evolution due to an optical homodyne measurement~\cite{gross2018qubit}. Specifically, the system in state $\hat{\rho}$ repeatedly and weakly interacts with an auxiliary qubit prepared in state $\ket{0}$ for a duration of $\Delta t$ via the interaction Hamiltonian  
	\begin{equation}\label{eq:hamint}
		\hat{H} = -i \sqrt{\frac{\gamma}{\Delta t}} ( \hat{A} \otimes  \hat{\sigma}^+ - \hat{A}^\dagger \otimes  \hat{\sigma}^-), 
	\end{equation}
	where $\hat{A}$ is the system operator being probed. Following each interaction, the auxiliary qubit is measured in the Pauli $X$ basis with outcomes $\{-1,+1\}$. The measurement projects this qubit into state $\ket{\pm}=\frac{1}{\sqrt{2}}(\ket{0}\pm\ket{1})$. The state of the system conditioned on the measurement outcome ($\pm1$) is then $\hat{\rho}+\Delta \hat{\rho}_\pm$.  The homodyne difference $\Delta\hat{\rho}^{(\rm{m})}_\pm$ up to order $\Delta \tau = \gamma \Delta t$  is given by
	\begin{align}
		\Delta \hat{\rho}^{(\rm{m})}_\pm &= \Delta I_h \mathcal{H}(\hat{A})[\hat{\rho}] +
		\Delta \tau \mathcal{D}(\hat{A})[\hat{\rho}],
	\end{align}
	where $\Delta I_h =( \pm\sqrt{\Delta \tau}-\Delta \tau{\rm{Tr}}[(\hat{A}+\hat{A}^\dagger)\hat{\rho}])$ is known as the innovation of the measurement, $\mathcal{H}(\hat A)[\hat{\rho}] = \hat A \hat{\rho} + \hat \rho \hat{A}^\dagger  - \hat \rho {\rm{Tr}}[(\hat{A}+\hat{A}^\dagger)\hat{\rho}]$ and $\mathcal{D}(\hat A)[\hat{\rho}] = \hat A \hat{\rho} \hat A^\dagger -\frac{1}{2} \{\hat A^\dagger \hat A,  \hat\rho\}$. We now feed forward the measurement results and apply a Hamiltonian operation 
	\begin{equation}\label{eq:hamff}
		\hat{H}_{\pm} =\pm \sqrt{\frac{\alpha}{\Delta t}} \hat{B},
	\end{equation}
	for the duration of $\Delta t$, where $\alpha$ signifies the strength of feed forward. The time evolution under this Hamiltonian up to order $\Delta \tau$ (assuming $\alpha$ and $\gamma$ are of the same order of magnitude) is given by 
	\begin{align}
		\hat U_\pm = \hat{I}\mp i\sqrt{\frac{\alpha}{\gamma} \Delta \tau} \hat B -\frac{1}{2}\frac{\alpha}{\gamma} \Delta \tau \hat B ^2. 
	\end{align}
	Therefore, the total change in the state of the system after a measurement and feed forward step up to order $\Delta \tau$ is
	\begin{align}
		\Delta \hat \rho^{\rm{(mff)}}_\pm &= \hat U_\pm (\hat \rho + \Delta \hat \rho^{(\rm{m})}_\pm)\hat U_\pm\\
		&= \Delta \hat \rho^{(\rm{m})}_\pm \mp i\sqrt{\frac{\alpha}{\gamma}\Delta \tau} [\hat{B},\hat{\rho}] + \frac{\alpha}{\gamma}\Delta \tau  \hat B \hat \rho \hat B -\frac{1}{2} \frac{\alpha}{\gamma} \Delta\tau\{\hat{B}^2 ,\hat\rho\}- i \sqrt{\frac{\alpha}{\gamma}} \hat{B} \Delta \tau  \mathcal{H}(\hat{A})[\hat{\rho}] + i\sqrt{\frac{\alpha}{\gamma}} \Delta \tau \mathcal{H}(\hat{A})[\hat{\rho}] \hat B\\
		&=\Delta \hat \rho^{(\rm{m})}_\pm \mp i\sqrt{\frac{\alpha}{\gamma}\Delta \tau} [\hat{B},\hat{\rho}]+\frac{\alpha}{\gamma}\Delta \tau \mathcal{D}( \hat B) [\hat\rho] - i \sqrt{\frac{\alpha}{\gamma}} \Delta \tau  [\hat{B},\mathcal{H}(\hat{A})[\hat{\rho}]]\\
		&=\Delta \hat \rho^{(\rm{m})}_\pm \mp i\sqrt{\frac{\alpha}{\gamma}\Delta \tau} [\hat{B},\hat{\rho}]+\frac{\alpha}{\gamma}\Delta \tau \mathcal{D}( \hat B) [\hat\rho] - i \sqrt{\frac{\alpha}{\gamma}} \Delta \tau  [\hat{B},\hat A \hat{\rho} + \hat \rho \hat{A}^\dagger  - \hat \rho {\rm{Tr}}[(\hat{A}+\hat{A}^\dagger)\hat{\rho}]].
	\end{align}
	Note that $\hat{A}$ is a Hermitian operator. Therefore, we use the fact that $\hat{A}=\hat{A}^\dagger$ to further simplify the expression for $\Delta \hat \rho^{\rm{(mff)}}_\pm$ and obtain
	\begin{align}
		\Delta \hat \rho^{\rm{(mff)}}_\pm &=\Delta \hat \rho^{(\rm{m})}_\pm \mp i\sqrt{\frac{\alpha}{\gamma}\Delta \tau} [\hat{B},\hat{\rho}]+\frac{\alpha}{\gamma}\Delta \tau \mathcal{D}( \hat B) [\hat\rho] - i \sqrt{\frac{\alpha}{\gamma}} \Delta \tau [\hat{B},\hat A \hat{\rho} + \hat \rho \hat{A}  - 2 \hat \rho {\rm{Tr}}(\hat{A}\hat{\rho})]\\
		& = \Delta \hat \rho^{(\rm{m})}_\pm -i\sqrt{\frac{\alpha}{\gamma}}(\pm \sqrt{\Delta \tau} -2 \Delta \tau {\rm{Tr}}(\hat{A}\hat{\rho}))[\hat{B},\hat{\rho}]+\frac{\alpha}{\gamma}\Delta \tau \mathcal{D}( \hat B) [\hat\rho] - i \sqrt{\frac{\alpha}{\gamma}} \Delta \tau [\hat{B},\hat A \hat{\rho} + \hat \rho \hat{A}]   \\
		& = \Delta \hat \rho^{(\rm{m})}_\pm -i\sqrt{\frac{\alpha}{\gamma}}\Delta I_h [\hat{B},\hat{\rho}]+\frac{\alpha}{\gamma}\Delta \tau \mathcal{D}( \hat B) [\hat\rho] - i \sqrt{\frac{\alpha}{\gamma}} \Delta \tau [\hat{B},\hat A \hat{\rho} + \hat \rho \hat{A}].
	\end{align}
	We now take the continuous limit where $\Delta I_h$ becomes $\sqrt{\gamma}dW$, a Wiener process satisfying $\expval{dW}=0$ and $\expval{dW^2}=dt$, and $\Delta \tau$ becomes $\gamma dt$ to find 
	\begin{align}
		d \hat \rho_c &= \sqrt{\gamma}dW \mathcal{H}(\hat{A})[\hat{\rho}] +   \gamma dt \mathcal{D}(\hat{A})[\hat{\rho}] -i\sqrt{\frac{\alpha}{\gamma}}\sqrt{\gamma}dW[\hat{B},\hat{\rho}]+\frac{\alpha}{\gamma}\gamma dt \mathcal{D}( \hat B) [\hat\rho] - i \sqrt{\frac{\alpha}{\gamma}} \gamma dt [\hat{B},\hat A \hat{\rho} + \hat \rho \hat{A}]
		\\
		&= \sqrt{\gamma}dW \mathcal{H}(\hat{A})[\hat{\rho}] +   \gamma dt \mathcal{D}(\hat{A})[\hat{\rho}] -i\sqrt{\alpha}dW[\hat{B},\hat{\rho}]+\alpha dt \mathcal{D}( \hat B) [\hat\rho] - i \sqrt{\alpha\gamma} dt [\hat{B},\hat A \hat{\rho} + \hat \rho \hat{A}]\\
		&= \sqrt{\gamma}dW \mathcal{H}(\hat{A})[\hat{\rho}]-i\sqrt{\alpha}dW[\hat{B},\hat{\rho}] +   dt \mathcal{D}(\sqrt{\gamma}\hat{A}-i\sqrt{\alpha}\hat{B})[\hat{\rho}]  - i \frac{\sqrt{\alpha\gamma}}{2} dt [\hat{A}\hat{B}+\hat{B}\hat{A}, \hat{\rho} ],
	\end{align}
	where $\hat \rho_c$ is the state conditioned on continuous measurement signal. Averaging over the measurement results, we find that the unconditional state evolves as 
	\begin{equation}
		\frac{d \hat \rho}{dt} = \mathcal{D}(\sqrt{\gamma}\hat{A}-i\sqrt{\alpha}\hat{B})[\hat{\rho}]  - i \frac{\sqrt{\alpha\gamma}}{2} [\hat{A}\hat{B}+\hat{B}\hat{A}, \hat{\rho} ].
	\end{equation}

	To illustrate these measurement and feedforward processes, we used a circuit representation in Fig. 1 in the main text. Below, we describe how the gate decomposition corresponding to the weak measurements and feedforward operations considered in this work is obtained.
	
	First, let us consider weak measurements of $\hat{A} =\frac{1}{\sqrt{n}} \sum_j u_{j} \hat{Z}_j$. This is realized by coupling the system to an auxiliary qubit with the Hamiltonian in Eq.~\eqref{eq:hamint}, evolving for time $\Delta t$, and measuring the auxiliary qubit in the Pauli $X$ basis. The Hamiltonian for this choice of $\hat{A}$ can be expressed as 
	\begin{align}
		\hat{H} &=  -i \sqrt{\frac{\gamma}{\Delta t}} ( \hat{A} \otimes  \hat{\sigma}^+ - \hat{A}^\dagger \otimes  \hat{\sigma}^-) \\ 
		&=   -i \sqrt{\frac{\gamma}{\Delta t}}  \left(\frac{1}{\sqrt{n}} \sum_j u_{j} \hat{Z}_j\right)\otimes(\hat{\sigma}^+-\hat{\sigma}^-)\\
		&=  \sqrt{\frac{\gamma}{\Delta t}}  \left(\frac{1}{\sqrt{n}} \sum_j u_{j} \hat{Z}_j\right)\otimes \hat{Y}.
	\end{align}
	Therefore, the unitary evolution $\hat{U}_A$ generated by this Hamiltonian after time $\Delta t$ is given by 
	\begin{align}
		\hat{U}_A &= \exp(-i\Delta t\sqrt{\frac{\gamma}{\Delta t}}  \left(\frac{1}{\sqrt{n}} \sum_j u_{j} \hat{Z}_j\right)\otimes \hat{Y}) \\ 
		&= \exp(-i\sqrt{\gamma\Delta t}  \left(\frac{1}{\sqrt{n}} \sum_j u_{j} \hat{Z}_j\right)\otimes \hat{Y}) \\
		&= \prod_j \exp(-i\sqrt{\frac{\gamma\Delta t}{n}}  u_{j} \hat{Z}_j\otimes \hat{Y})\\
		&= \prod_j \hat{R}^{(j)}_{YZ}(\theta u_{j}),
	\end{align}
	where we introduced $\theta=\sqrt{\gamma \Delta t/n}$ and used $\hat{R}^{(j)}_{YZ}(\theta u_j)$ to denote a rotation of angle $\theta u_j$ around the $YZ$ axis of the auxiliary qubit and the $j$th system qubit.
	
	Next, we consider the gate implementation of the feedforward operation generated by the Hamiltonian in Eq.~\eqref{eq:hamff}. Let $\hat{B} =\frac{1}{\sqrt{n}} \sum_j v_{j} \hat{Z}_j$. The corresponding evolution operator  for time $\Delta t$ on system qubits, given the measurement outcome $\pm1$ on the auxiliary qubits is  given by 
	\begin{align}
		\hat{U}_\pm &= \exp\left(\mp i \Delta t \sqrt{\frac{\alpha}{\Delta t}}\hat B \right) \\
		&=\exp\left(\mp i  \sqrt{\frac{\alpha \Delta t}{n}}\sum_j \hat v_{j} \hat{Z}_j \right) \\
		&=\prod_j \exp\left(\mp i  \sqrt{\frac{\alpha \Delta t}{n}} \hat v_{j} \hat{Z}_j \right)\\
		&= \prod_j  \hat{R}^{(j)}_Z (\pm \theta v_j),
	\end{align}
	where we introduced $\hat{R}^{(j)}_Z (\theta v_j)$ to denote a rotation of $\theta v_j$ around the $Z$ axis on qubit $j$, and assumed $\alpha$ to be the same as $\gamma$ for simplicity.
	
	Therefore, we can achieve the desired measurement and feedforward operations using $\hat{R}_{YZ}$ and classically controlled $\hat{R}_Z$ gates with $\theta\ll 1$, along with mid-circuit measurements of the auxiliary qubit.